# Vacuum Rabi splitting and intracavity dark state in a cavity-atoms system

Gessler Hernandez, Jiepeng Zhang, and Yifu Zhu

Department of Physics, Florida International University, Miami, Florida 33199

Abstract

We report experimental measurements of the transmission spectrum of an optical cavity coupled with cold Rb atoms. We observe the multi-atom vacuum Rabi splitting of a composite cavity and atom system. When a coupling field is applied to the atoms and induces the resonant two-photon Raman transition with the cavity field in a $\Lambda$ type system, we observe a cavity transmission spectrum with two vacuum Rabi sidebands and a central peak representing the intracavity dark state. The central peak linewidth is significantly narrowed by the dark-state resonance and its position is insensitive to the frequency change of the empty cavity.

Cavity QED studies interactions of atoms and electromagnetic modes of an optical cavity. The most fundamental system consists of a single two-level atom coupled to a single cavity mode [1]. It has been shown that the composite atom-cavity system exhibits a double-peaked transmission spectrum with the peak separation referred to as the vacuum Rabi splitting and determined by the atom-cavity coupling coefficient $g = \mu\sqrt{\omega_a/2\hbar\varepsilon_0 V}$ [2-3]. To observe the vacuum Rabi splitting in the optical wavelength range, it needs to have a g value greater or comparable with the decay rates of the cavity and the atomic system, which requires a high finesse cavity with a small mode volume [4]. However, if N atoms collectively interact with the cavity mode, the coupling coefficient becomes $g = \mu\sqrt{N\omega_a/2\hbar\varepsilon_0 V}$ and the multi-atom vacuum splitting may then be observed in a cavity with a moderate mode volume and finesse [5-7]. Studies of atom-cavity interactions can be extended to a composite system of an optical cavity and coherently prepared multi-level atoms, in which the atomic coherence and interference in coherently prepared atoms may be enhanced. For example, electromagnetically induced transparency (EIT)/Coherent population trapping (CPT) can be produced in a three-level Λ-type system [8-9], and its manifestation in a cavity-atom system may be useful for a variety of fundamental studies and practical applications. It has been shown that intracavity EIT results in an ultra narrow spectral linewidth, which may be used for frequency stabilization and high-resolution spectroscopic measurements [10]. Experimental studies of coherently prepared hot atoms confined in a cavity have observed the EIT linewidth narrowing [11] and other interesting phenomena such as control of optical multistability [12].

Here we report an experimental study of a cavity-atom composite system consisting of cold Rb atoms confined in the mode volume of a 5 cm long, near confocal optical cavity. The cold atoms are coupled with the cavity mode and an additional coupling laser, which forms a three-level Λ configuration. We measure the cavity transmission spectrum of a weak probe laser mode-matched to the optical cavity. When the coupling laser is absent, we observe the multi-atom vacuum Rabi splitting. When the coupling laser is turned on and induces a resonant two-photon Raman transition in the composite atom-cavity system, we observe a three-peaked spectrum consisting of two broad sidebands representing the vacuum Rabi splitting and a narrow central peak manifested by the dark-state resonance of the two-photon Raman transition. The central peak has a linewidth smaller than the natural linewidth and the cavity linewidth, and is limited by the laser linewidth in our experiment. The transmission frequency of the central peak is determined by the two-photon Raman resonance and is insensitive to the change of the empty cavity frequency. The experimental results agree with theoretical calculations based on a semiclassical analysis.

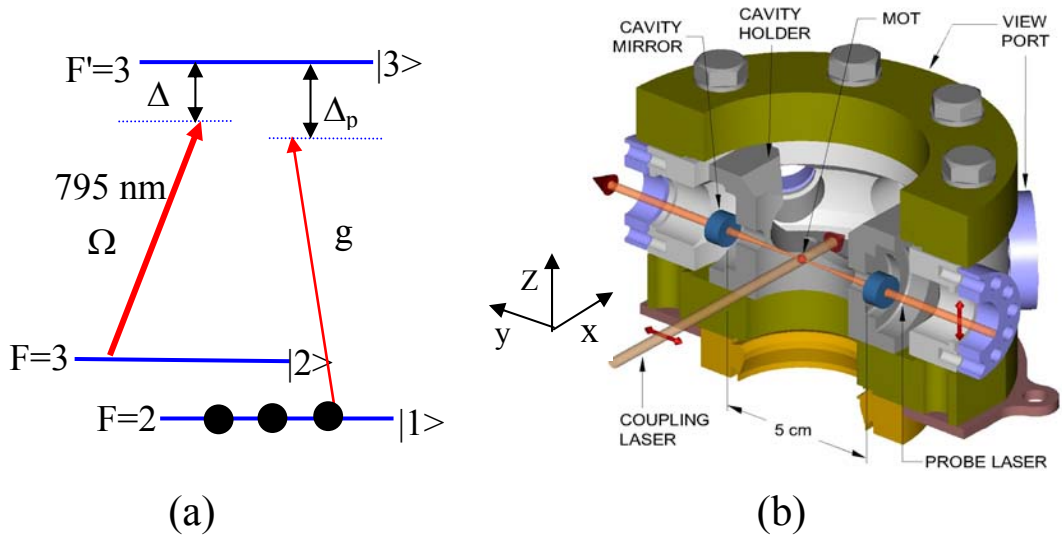

(a)                          (b)

Fig. 1 (a) $^{85}$Rb atoms interacting with a coupling field and a cavity field, which forms a three-level Λ-type system. The spontaneous decay rate of the excited state |3> is $\Gamma$ (=2π×5.4×10$^6$ s$^{-1}$). (b) Schematic drawing of the cavity apparatus. The coupling (probe) laser is linearly polarized along the y (z) direction.

The composite atom-cavity system in our experiment can be viewed as a coupled Λ system shown in Fig. 1a. The standing-wave cavity mode couples the atomic transition |1>-|3> (the $^{85}$Rb D1 F=2-F'=3 transition). The coupling field drives the atomic transition |2>-|3> (the $^{85}$Rb D1 F=3-F'=3 transition) with Rabi frequency 2Ω and forms a standard Λ-type configuration. $\Delta = \nu - \nu_{32}$ is the coupling laser-atom detuning, $\Delta_c = \nu_c - \nu_{31}$ is the cavity mode-atom detuning, and $\Delta_p = \nu_p - \nu_{31}$ is the probe laser-atom detuning. Under the condition of the two-photon Raman resonance ($\Delta = \Delta_p$), a dark state is created, which leads to the suppressed light absorption and rapidly varying dispersion near the two-photon resonance [8-9]. When $\Delta = \Delta_p = 0$, the dark state resonance corresponds to the resonant EIT, under which it has been shown that the cavity transmission peak occurs at the resonant frequency [10] $\nu_r = \frac{\nu_c}{1+\eta} + \frac{\eta \nu_p}{1+\eta}$. Here $\nu_p$ is the probe laser frequency $\nu_p = \nu + \nu_{21}$ ($\nu_{21}$ is the frequency separation between the two ground states $|1\rangle$ and $|2\rangle$) and $\eta = \frac{\nu_r \ell}{2L} \frac{\partial \chi'}{\partial \nu_p}$ is a coefficient characterizing the dispersion change near the two-photon Raman resonance ($\ell$ is the medium length, L is the cavity length, and $\chi'$ is the real part of the medium susceptibility at the probe frequency). The linewidth of the cavity transmission peak is $\Delta \nu = \frac{1 - R\kappa}{\sqrt{\kappa(1-R)}} \frac{C}{1+\eta}$ [10]. Here C is the empty cavity linewidth, R is the intensity reflectivity of the cavity mirror, and $\kappa = \exp(-2\pi \nu_p \chi'' \ell / c)$ is the single pass medium absorption ($\chi''$ is the imaginary part of the medium susceptibility at the probe frequency). When $\eta \gg 1$, the linewidth of the cavity transmission peak is reduced by a factor of $\eta$ relative to the empty cavity linewidth and is ultimately limited by the ground state decoherence rate γ that can be many orders less than the atomic natural linewidth Γ. Also the cavity resonant frequency is essentially independent of the empty cavity frequency and nearly equal to the probe frequency $\nu_p$.

The experiment is done with cold $^{85}$Rb atoms confined in a magneto-optical trap (MOT) produced at the center of a 10-ports stainless-steel vacuum chamber. The MOT is obtained with two extended-cavity diode lasers with a beam diameter of ~ 1 cm: one with output power ~ 30 mW is used as the cooling and trapping laser supplying six perpendicular retro-reflected beams, and another with output power ~15 mW is used as the repump laser. The trapped $^{85}$Rb atom cloud is ~ 1.5 mm in diameter and the measured optical depth $n\sigma_{13}\ell$ is ~ 3. A schematic diagram of the cavity apparatus is depicted in Fig. 1(b). The standing-wave cavity consists of two mirrors of 5 cm curvature with a mirror separation of ~ 5 cm and is mounted on an Invar holder enclosed in the vacuum chamber. The empty cavity finesse is measured to be ~ 200. Movable anti-Helmholtz coils are used so the MOT position can be finely adjusted to coincide with the cavity center. A third extended-cavity diode laser with a beam diameter ~ 5 mm and output power ~ 35 mW is used as the coupling laser that propagates along the x direction and directed to overlap with the MOT through a vacuum viewport. A fourth extended-cavity diode laser is used as the probe laser, which is attenuated and split into two parts. One part propagates in the y direction and is coupled into the cavity. The transmitted cavity light passes through an iris and then is collected by a photodiode. Another part of the probe beam propagates nearly parallel to the coupling laser (at an angle of ~3°) in the same x-y plane, overlaps with the MOT from free space, and is then collected by another photodiode, which provides the probe absorption spectrum in free space and serves as a reference for comparison with the recorded cavity transmission spectrum.

The experiment is run in a sequential mode with a repetition rate of 10 Hz. all lasers are turned on or off by acousto-optic modulators (AOM) according to the time sequence described below. For each period of 100 ms, ~99 ms is used for cooling and trapping of the $^{85}$Rb atoms, during which the trapping laser and the repump laser are turned on by two AOMs while the coupling laser and the probe laser are off. The time for the data collection lasts ~ 1 ms, during which the repump laser is turned off first, and after a delay of ~0.15 ms, the trapping laser and the current to the anti-Helmholtz coils of the MOT are turned off, and the coupling laser and the probe laser are turned on. After the coupling laser and probe laser are turned on by the AOMs for 0.1 ms, the probe laser frequency is scanned across the $^{85}$Rb $D_1$ F=2→F=3 transitions and the cavity transmission of the probe laser is then recorded versus the probe frequency detuning.

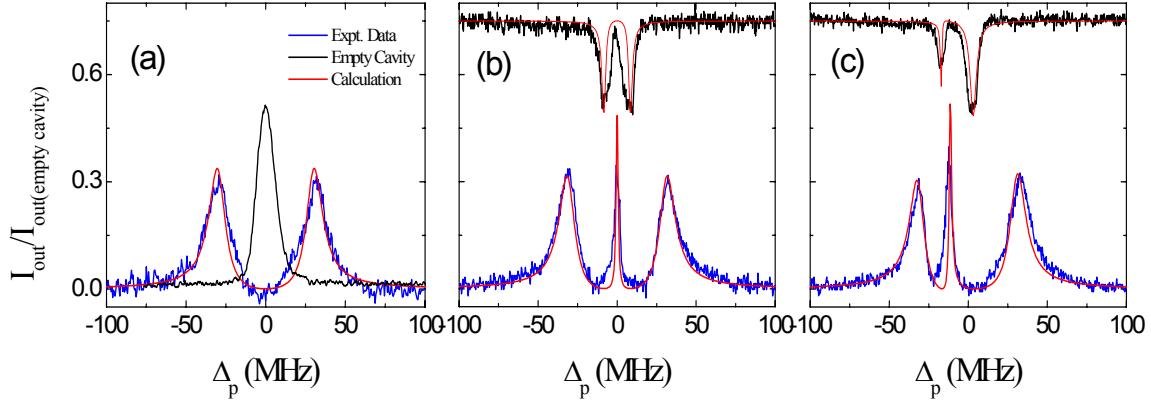

Fig. 2 Cavity transmission versus the probe detuning $\Delta_p$. Blue (red) lines are experimental data (calculations). The cavity detuning $\Delta_c=0$. (a) $\Omega=0$. (b) $\Omega\approx 8$ MHz and $\Delta\approx 0$. (c) $\Omega\approx 8$ MHz and $\Delta\approx -12$ MHz. For comparison, the free-space probe spectrum (in an arbitrary scale) is plotted in the top of (b) and (c), in which black (red) lines are experimental data (calculations). The other parameters used in the calculations are $n\sigma_{13}\ell=2.5$, R=0.97, and the ground state decoherence rate $\gamma=0.02\ \Gamma$.

Fig. 2 plots the measured cavity transmission of the probe laser (normalized to the peak intensity of the empty-cavity transmission $I_{out(empty\ cavity)}$) versus the probe frequency detuning $\Delta_p$ (the empty cavity frequency is tuned to $\nu_c=\nu_{31}$). Blue lines are experimental data and red lines are calculations based on a semiclassical analysis. For reference, the transmission peak of the empty cavity (no atoms in the cavity) is plotted as the dark line in Fig. 2(a) and its vertical scale is reduced by 2 times. When the coupling laser is turned off ($\Omega=0$), the cold Rb atoms can be viewed as a two-level system and the cavity transmission is plotted as the blue line in Fig. 2(a). We observe two transmission peaks at $\Delta_p=\pm g$. The peak separation represents the multi-atom vacuum Rabi splitting (the normal cavity-atom modes) and is given by $2g=2\mu\sqrt{N\omega_a/2\hbar\varepsilon_0 V} \approx 60$ MHz, from which we derive that there are ~ $10^4$ atoms in the cavity mode volume of V~$7\times 10^{-4}$ cm$^3$. The measurements were taken at the low power levels of the input probe laser at which the transmitted power of the empty cavity (without atoms) $\leq 0.1$ μW. We observed that at higher probe powers, the line shape of the vacuum Rabi splitting become asymmetrical due to saturation of the intra-cavity field as reported in ref. [7]. When the coupling laser is turned on, the cavity transmission spectrum versus $\Delta_p$ exhibits three peaks: the two vacuum Rabi sidebands at $\Delta_p \approx \pm g$ and a central peak at $\Delta_p=\Delta$ that represents the dark-state resonance and is narrowed by the dark-state manifested frequency pulling and absorption suppression. The central peak in Fig. 2(b) corresponds to the EIT resonance ($\Delta_p=\Delta=0$). Our measurements show that when $\Delta \neq 0$, the intra-cavity dark state is shifted accordingly and always occurs at the two-photon Raman resonance $\Delta_p=\Delta$. Fig 2(c) shows the measured spectrum for $\Delta \approx -12$ MHz, in which the cavity dark-state resonance occurs at $\Delta_p=-12$ MHz. For comparison, the probe absorption spectra under these conditions in free space are plotted in the top of Fig. 2(b) and 2(c), and show the resonant EIT ($\Delta=0$, Fig.2(b)) and the off-resonant spectral features ($\Delta \approx -12$ MHz, Fig.2(c)).

Our results show that with EIT, the splitting of the two vacuum Rabi sidebands becomes $2\sqrt{g^2+\Omega^2}$. In our experiment, g≈30 MHz and Ω≈8 MHz, the EIT and the dark state resonance results in the narrow central peak at $\Delta_p=0$ for the composite cavity-atom system, but its effect on the vacuum Rabi sidebands is small (the separation of the two vacuum Rabi sidebands in Fig. (2b) differs from that of Fig. (2a) by ~1 MHz). When the coupling laser is detuned from the atomic resonance ($\Delta \neq 0$), the measurements show that the frequency pulling induced by the atomic dispersion near the two-photon Raman resonance shifts the cavity resonant frequency $\nu_r$ such that the narrow cavity transmission occurs at the resonant two-photon transition $\Delta_p=\Delta \neq 0$ where the intracavity dark state is formed. The linewidth (FWHM) of the intra-cavity dark state is measured to be ~ 1.2 (±0.4) MHz, which is limited by the laser linewidth (~1 MHz) in our experiment, and is smaller than the Rb natural linewidth (Γ=5.6 MHz) and the empty cavity linewidth (~14 MHz). The experimental measurements agree with the theoretical calculations and the value of $\eta$ is derived to be ~ 25.

Fig. 3 plots the measured cavity transmission of the probe laser (normalized to the peak intensity of the empty-cavity transmission $I_{out(empty\ cavity)}$) versus the probe frequency detuning $\Delta_p$ when the empty cavity frequency $\nu_c$ is detuned from the atomic transition frequency $\nu_{31}$ (the coupling frequency is kept on resonance (Δ=0)). It shows that when the cavity is detuned ($\nu_c \neq \nu_{31}$), the vacuum Rabi peaks, representing the normal modes of the coupled cavity-atoms system, shift in their positions and have unequal amplitudes, but the narrow central peak is always located near $\Delta_p$=Δ=0 where the two-photon Raman transition |1>-|3>-|2> is resonant. That is, the intra-cavity dark state is induced by the two-photon Raman resonance, its position is given by $\Delta_p$=Δ and is insensitive to the change of the empty cavity frequency.

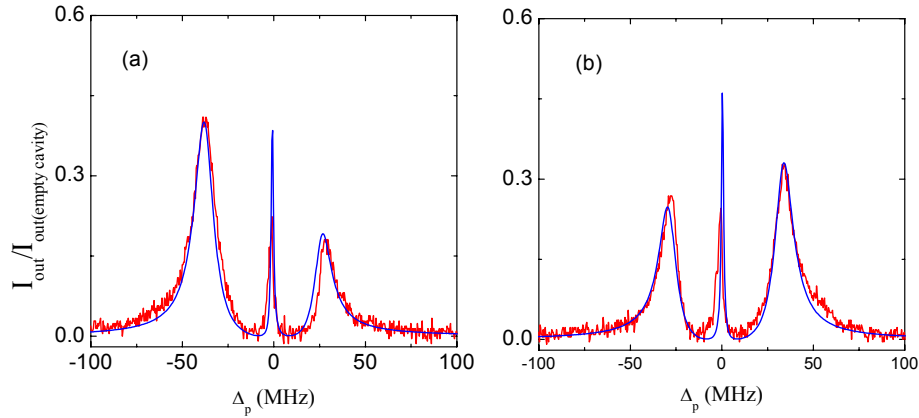

Fig. 3 Cavity transmission versus the probe detuning $\Delta_p$. Blue (red) lines are experimental data (calculations). The coupling detuning Δ=0. (a) Ω≈8 MHz and $\Delta_c$≈13 MHz. (b) Ω≈8 MHz and $\Delta_c$≈-5 MHz. The other parameters are the same as those in Fig. 2.

Under the condition of the two-photon Raman resonance ($\Delta_p=\Delta$), we plot the position of the intra-cavity dark state (the frequency of the central transmission peak) versus the coupling frequency detuning $\Delta$ in Fig.

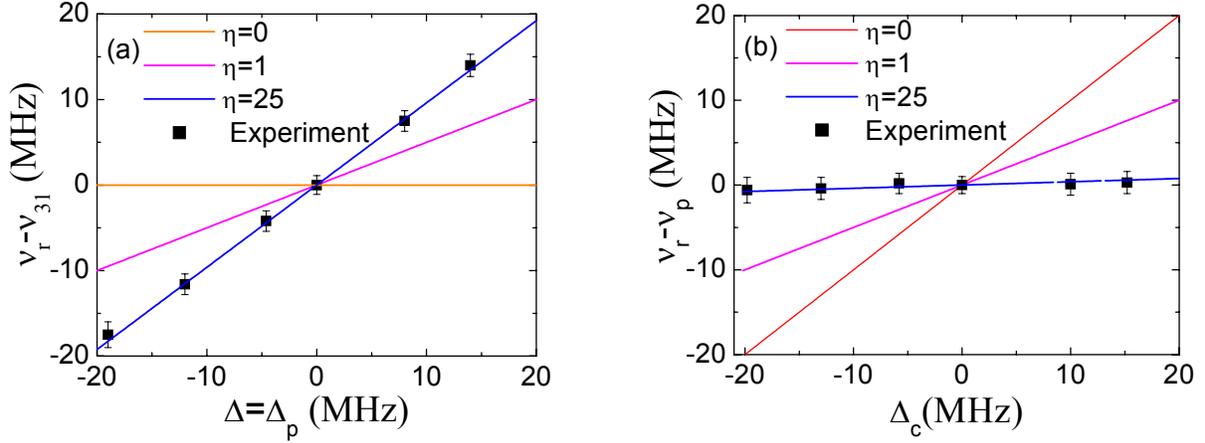

Fig. 4 Resonant frequency of the intra-cavity dark state versus (a) the frequency detuning $\Delta=\Delta_p$ and (b) the frequency detuning $\Delta_c$ of the empty cavity. Lines are the calculations and dots are experimental data. For (a), the empty cavity frequency is tuned to the atomic transition frequency $\nu_{31}$ ($\Delta_c=0$). For (b), the probe and the coupling detunings are kept at $\Delta_p=\Delta=0$.

Fig. 4(a) and also versus the empty cavity detuning $\Delta_c$ in Fig. 4(b). The experimental data are plotted as solid dots and the calculated results for several $\eta$ values are plotted in solid lines. The experimental measurements demonstrate that the cavity resonant frequency $\nu_r$ (the intracavity dark-state resonance) is nearly equal to the probe frequency $\nu_p$ at the two-photon Raman resonance (Fig. 4(a)) and is insensitive to the frequency change of the empty cavity (Fig. 4(b)). Such characteristic of the coupled cavity-atom system is useful for the frequency standard and atomic clocks based on EIT/ CPT [13]. There are two advantages of a cavity-atom composite system for such applications: first, a strong coupling field can be used to create a large EIT/CPT window, which increases the signal to noise ratio but does not cause the power broadening of the cavity transmission; second, the cavity transmission peak is always kept on the two-photon Raman resonance $\Delta_p=\Delta$ and is insensitive to changes of the empty cavity frequency caused by the cavity length drift due to the thermal and mechanical instabilities.

In conclusion, we have measured the transmission spectrum of an optical cavity coupled with coherently prepared cold Rb atoms. When the intra-cavity dark state is induced by a coupling laser in the cavity and $\Lambda$-type atomic system, the cavity transmission spectrum exhibits three peaks: two sidebands associated with the multi-atom vacuum Rabi splitting and a narrow central peak representing the cavity dark-state resonance. The experimental results agree with the theoretical results of ref. [10] for the resonant EIT at $\Delta_p=\Delta=0$ and also show that they can be extended to the detuned $\Lambda$ system without EIT as long as the two-photon Raman resonance is satisfied. The cavity-atom composite system can be used for a variety of fundamental studies [14] and may be useful for the EIT/CPT based precision measurement applications [13,15].

This work is supported by the National Science Foundation under Grant No. 0456766.

References

1. E. T. Jaynes and F. W. Cummings, Proc. IEEE 51, 89(1963).

2. J. J. Sanchez-Mondragon, N. B. Narozhny, and J. H. Eberly, Phys. Rev. Lett. 51, 550(1983).

3. Cavity Quantum Electrodynamics, edited by P. R. berman (Academic, San Diego, 1994).

4. A. Boca, R. Miller, K. M. Birnbaum, A. D. Boozer, J. McKeever, and H. J. Kimble, Phys. Rev. Lett. 93, 233603 (2004).

5. G. S. Agarwal, Phys. Rev. Lett. 53, 1732(1984).

6. Y. Zhu, D. J. Gauthier, S. E. Morin, Q. Wu, H. J. Carmichael, and T. W. Mossberg, Phys. Rev. Lett. **64**, 2499 (1990).

7. J. Gripp, S. L. Mielke, and L. A. Orozco, Phys. Rev. A 56, 3262(1997).

8. S. E. Harris, Phys. Today 50, 36 (1997).

9. E. Arimondo, in *Progress in Optics* (E. Wolf ed.) Vol. 31, 257(1996).

10. M. D. Lukin, M. Fleischhauer, M. O. Scully, and V. L. Velichansky, Opt. Lett. **23**, 295 (1998).

11. H. Wang, D. J. Goorskey, W. H. Burkett, and M. Xiao, Opt. Lett. **25**, 1732 (2000).

12. A. Joshi and M. Xiao, Phys. Rev. Lett. 91, 143904 (2003).

13. S. Knappe, V. Shah, P. Schwindt, L. Hollberg L, J. Kitching , L. A. Liew , J. Moreland, Appl. Phys. Lett. 85, 1460 (2004).

14. J. Ye and T. W. Lynn, in *Advances in Atomic, Molecular, and Optical Physics*, edit. by B. Bederson and H. Walther, Vol. 49, p1 (2003).

15. P. Schwindt, S. Knappe S, V. Shah, L. Hollberg, J. Kitching, L. A. Liew, and J. Moreland, Appl. Phys. Lett. 85, 6409(2004).